\documentclass[%
 reprint,
 amsmath,amssymb,
 aps,
prl,
]{revtex4-2}

\usepackage{graphicx}% Include figure files
\usepackage{dcolumn}% Align table columns on decimal point
\usepackage{bm}% bold math
\usepackage{amsfonts,mathrsfs,textcomp,dsfont,esint,array,multirow,url}
\usepackage[mathscr]{euscript}

\def\CA{{\cal A}}
\def\CF{{\cal F}}
\def\CH{{\cal H}}\def\CI{{\cal I}}
\def\CM{{\cal M}}
\def\CN{{\cal N}}

\def\CT{{\cal T}}
\def\CW{{\cal W}}

\def\a{\alpha}\def\b{\beta}

\newcommand{\be}{\begin{equation}}
\newcommand{\ee}{\end{equation}}
\newcommand{\bea}{\begin{equation} \begin{aligned}} \newcommand{\eea}{\end{aligned} \end{equation}}
\newcommand{\ba}{\begin{array}}

\newcommand{\smashlastub}[1]{%
  \sbox0{\let\smash\relax$#1$}
  \vphantom{\usebox0}
  \sbox2{$#1$}
  \raisebox{\dimexpr(\ht0-\ht2)}{\usebox2}
}

\begin{document}

\preprint{APS/123-QED}

\title{Three-Dimensional Topological Field Theories\\ and Non-Unitary Minimal Models}

\author{Dongmin Gang}
\email{ arima275@snu.ac.kr}
 \affiliation{Department of Physics and Astronomy \& Center for Theoretical Physics,
Seoul National University, 1 Gwanak-ro, Seoul 08826, Korea}%Lines break automatically or can be forced with \\
\author{Heeyeon Kim}%
 \email{heeyeon.kim@kaist.ac.kr}
\affiliation{%
Department of Physics, Korea Advanced Institute of Science and Technology,
Daejeon 34141, Republic of Korea
}%

\author{Spencer Stubbs}
 \email{spencer.stubbs@physics.rutgers.edu}
\affiliation{
NHETC and Department of Physics and Astronomy, Rutgers University, 126 Frelinghuysen Rd.,
Piscataway NJ 08855, USA
}%


\begin{abstract}
We find an intriguing relation between a  class of 3-dimensional non-unitary topological field theories (TFTs) and Virasoro minimal models $M(2,2r+3)$ with $r\geq 1$. The TFTs are constructed by topologically twisting 3d $\CN=4$ superconformal field theories (SCFTs) of rank-0, i.e. having zero-dimensional Coulomb and Higgs branches. We present ultraviolet (UV) field theory descriptions of the SCFTs with manifest $\CN=2$ supersymmetry, which we argue is enhanced to $\CN=4$ in the infrared. From the UV description, we compute various partition functions of the TFTs and reproduce some basic properties of the minimal models, such as their characters and modular matrices. We expect more general correspondence between topologically twisted 3d $\CN=4$ rank-0 SCFTs and 2d non-unitary rational conformal field theories.  
\end{abstract}

\maketitle

%\tableofcontents

\section{\label{sec:intro}Introduction}

% This sample document demonstrates proper use of REV\TeX~4.2 (and
% \LaTeXe) in mansucripts prepared for submission to APS
% journals. Further information can be found in the REV\TeX~4.2
% documentation included in the distribution or available at
% \url{http://journals.aps.org/revtex/}.

% When commands are referred to in this example file, they are always
% shown with their required arguments, using normal \TeX{} format. In
% this format, \verb+#1+, \verb+#2+, etc. stand for required
% author-supplied arguments to commands. For example, in
% \verb+\section{#1}+ the \verb+#1+ stands for the title text of the
% author's section heading, and in \verb+\title{#1}+ the \verb+#1+
% stands for the title text of the paper.

% Line breaks in section headings at all levels can be introduced using
% \textbackslash\textbackslash. A blank input line tells \TeX\ that the
% paragraph has ended. Note that top-level section headings are
% automatically uppercased. If a specific letter or word should appear in
% lowercase instead, you must escape it using \verb+\lowercase{#1}+ as
% in the word ``via'' above.

Two-dimensional rational conformal field theory (RCFT) has played an essential role in many different problems in theoretical physics and mathematics. They are characterized by the property that the Hilbert space decomposes into a finite sum of the representations $V_\alpha$ and $V_{\bar \alpha}$ of some chiral algebras $\CA$ and $\bar{\CA}$,
\be
\CH = \bigoplus_{\alpha,\bar \alpha}\CM_{\alpha,\bar \alpha} V_\alpha \otimes V_{\bar \alpha}\ . 
\ee
The simplest class of RCFTs is the Virasoro minimal models, which have a finite number of irreducible representations of the Virasoro algebra. They have a wide range of applications in the study of two-dimensional critical systems, even when they are non-unitary.

The partition function of an RCFT on a torus with complex structure $\tau$ can be written as a combination of a finite number of holomorphic and anti-holomorphic functions in $q = e^{2\pi i\tau}$,
\be
Z(\tau,\bar\tau) = \sum_{\alpha,\bar \alpha} \CM_{\alpha,\bar \alpha}\chi_\alpha(q)\bar\chi_{\bar \alpha} (\bar q)\ ,
\ee
where the holomorphic functions $\chi_\alpha(q)$ are called the characters of the representations $V_\a$.  The invariance of the partition function under the modular transformation, 
\be\label{modular transformation}
\tau \rightarrow \frac{a\tau + b}{c\tau + d} \quad \text{ for } \left(\begin{array}{cc} a & b \\ c & d \end{array}\right) \in SL(2,\mathbb{Z})\ ,
\ee 
implies that the RCFT characters transform as vector-valued modular functions.

In this letter, we construct a novel class of three-dimensional topological field theories (TFTs) 
which are expected to support two-dimensional rational chiral algebras on their boundaries.
%which have boundary algebras whose characters correspond to those of non-unitary rational conformal field theories.
 We will argue that these 3d TFTs can be constructed from a certain family of 3d $\CN=4$ superconformal field theories (SCFTs). A key characteristic of these SCFTs is that they are rank-0, i.e. their Coulomb and Higgs branches are zero-dimensional. The first examples of such theories were discovered in \cite{Gang:2018huc,Gang:2021hrd}. 
 
 In general, these 3d theories do not admit a Lagrangian description that preserves the full $\CN=4$ symmetry. Instead, we will present an ultraviolet (UV) field theory description with manifest $\CN=2$ supersymmetry which flows to an infrared (IR) fixed point with enhanced supersymmetry. Each $\CN=4$ theory at the fixed point admits two topological twists \cite{Witten:1988ze} which produce two distinct 3d TFTs. These topological theories are in general non-unitary and do not have local operators.

 Despite the absence of a Lagrangian description of the IR theory, the $\CN=2$ UV description enables exact computations of various observables in the topologically twisted theories. These computations allow us to extract the data of the corresponding boundary algebra, such as its characters and modular data.

For concreteness, in this letter, we focus on a simple class of TFTs that reproduce the data of non-unitary Virasoro minimal models $M(2,2r+3)$ for $r \geq 1$. However, we expect that this correspondence exists for a more general class of rank-0 theories and non-unitary RCFTs. We will discuss their construction and classification in an upcoming paper \cite{GKS}.

\section{A class of 3d $\CN=4$ rank-0 theories}

\subsection{An $\CN=2$ description}
Let us consider the following class of 3d $\CN=2$ abelian Chern-Simons matter theories, which we call $\CT_r$:
\be\label{Tr UV}
U(1)^r_{K_r} ~ + ~ \Phi_{a=1,\cdots r}\ ,
\ee
with the superpotential deformation,
\be
\CW =V_{\mathbf{m}_1} +\ldots +V_{\mathbf{m}_{r-1}} \ .
\ee
The charge of the $a$-th chiral multiplet $\Phi_a$ under the $b$-th $U(1)$ gauge symmetry is $\delta_{ab}$.
%.
%
There are mixed Chern-Simons interactions among the abelian gauge fields given by the following level matrix \footnote{In the terminology of \cite{Closset:2018ghr}, $K_r$ is the bare CS level in the ``$U(1)_{-1/2}$ quantization''. The UV effective Chern-Simons level is $K_r - \frac{1}2 I$.}:
\begin{align}
\begin{split}
&K_r = 2	\begin{pmatrix}
	1 & 1 & 1 & \cdots & 1 & 1 \\
	1 & 2 & 2 & \cdots & 2 & 2 \\
	1 & 2 & 3 & \cdots & 3 & 3 \\
	\vdots & \vdots & \vdots & \ddots & \vdots & \vdots \\
	1 & 2 & 3 & \cdots & r-1 & r-1\\
	1 & 2 & 3 & \cdots & r-1 & r \\
\end{pmatrix}\; , \label{level}
\end{split}
\end{align}
which coincides with $2 C(T_r)^{-1}$, where $C(T_r)$ is the Cartan matrix of the tadpole graph, obtained by folding the Cartan matrix of $A_r$ in half. The $V_{\mathbf{m}_i}$'s are 1/2 BPS, gauge-invariant, bare monopole operators with fluxes \footnote{The gauge charge of the monopole operator under the $a$-th $U(1)$ gauge group is $ \sum_b  (K_r)_{ab}  m_b - \frac{1}{2} (|m_a|+m_a)$. %and one can confirm the gauge-invariance.
}:
\begin{align}
\begin{split}
&\mathbf{m}_1 =(2,-1,0,\ldots 0 ) \;,
\\
 &\mathbf{m}_2 =(-1,2,-1,0,\ldots 0 ) \;,
 \\
 &\ \vdots \;
 \\
 &\mathbf{m}_{r-1}=(0,\ldots, -1,2,-1)\;. 
 \end{split}
\end{align}

After the monopole deformation, the 3d $\CN=2$ gauge theory has an unbroken $U(1)$ flavor symmetry which we denote by $U(1)_A$. The charge $A$ of this flavor symmetry is
\begin{align}
A = \sum_{a=1}^r a M_a\;,
\end{align}
where $M_a$ is the topological charge of $a$-th $U(1)$ gauge symmetry.
The theory also has a $U(1)_R$ R-symmetry which can be mixed with the $U(1)_A$ flavor symmetry.  We denote the R-charge at general mixing parameter $\nu \in \mathbb{R}$ by $R_\nu$, i.e.,
\begin{align}
R_\nu = R_{0 } + \nu A\;.
\end{align}
We choose the reference R-charge, $R_0$, to be the superconformal R-charge, which can be determined by F-maximization \cite{Jafferis:2010un}.

\subsection{Supersymmetry enhancement}

Here we  claim that the $\CN=2$ gauge theory $\CT_r$ flows to an $\CN=4$ rank-0 SCFT in the IR with an accidental supersymmetry (SUSY) enhancement. For the $r=1$ case, SUSY enhancement was claimed in \cite{Gang:2018huc} by demonstrating several pieces of non-trivial evidence. We give similar evidence for general $r$. 
Under the SUSY enhancement, the manifest $U(1)_R \times U(1)_A$ symmetry is expected to become an $SO(4)_R \simeq SU(2)_C \times SU(2)_H$ R-symmetry with the following embedding
\begin{align}
R_{\nu} = (J_3^C+J_3^H) + \nu (J_3^C-J_3^H)\;.
\end{align}
Here $J^{C/H}_3$ are the Cartan generators of the $SU(2)_{C/H}$ R-symmetries, whose charges take half-integral values.

To see the SUSY enhancement, we compute the superconformal index $\CI_{\rm sci}(q,\eta;\nu)$ which is defined as
\begin{align}
\CI_{\rm sci} (q, \eta ;\nu ) :=  \textrm{Tr}_{\CH_{\rm rad}(S^2)} (-1)^{R_\nu} q^{\frac{R_\nu}2 + j_3} \eta^A\;.
\end{align}
Here $\CH_{\rm rad}(S^2)$ is the Hilbert space of radially quantized theory on $S^2$ 
and $j_3 \in \frac{\mathbb{Z}}2$ is the spin. The index can be computed via supersymmetric localization \cite{Kim:2009wb,Imamura:2011su} and we find 
\begin{align}
\CI_{\rm sci} (q, \eta \;\nu=0) =1-q -\left(\eta +\frac{1}\eta \right)q^{3/2}+O(q^2)\;.
\end{align}
Only $q^{\frac{1}2 \mathbb{Z}_{\geq 0}}$-terms appear in the index, which is the first sign of an enhancement.  Further, the terms $-\left(\eta +\frac{1}\eta \right)q^{3/2}$ can only come either from extra SUSY-current multiplets or chiral primary multiplets with superconformal R-charge $3$ \cite{Cordova:2016emh}.  Performing a semi-classical analysis of $\CH_{\rm rad}(S^2)$, one can verify that there are two 1/4 BPS dressed monopole operators  which have $R_0=j_3=1$ and $A=\pm 1$, which are exactly the same as that of 1/4 BPS operators in extra-SUSY multiplets.
The  monopole operators  are
\begin{align}
\begin{split}
&  \psi_r^* V_{\mathbf{m}}\; \textrm{with }  \mathbf{m}= \begin{cases}
(1)\;, \quad r=1
\\
(\mathbf{0}_{r-2},-1,1) \;, \quad r>1
\end{cases}
 \\
 & \textrm{and }\phi_1^2 \phi_2^2 \ldots \phi_r^2 V_{\mathbf{m}=(-1,\mathbf{0}_{r-1})}\;. \nonumber
 \end{split}
\end{align}
Here   $(\phi_a, \psi_a)$ are the (scalar, spinor)  in the $a$-th chiral multiplet.  
On the other hand, we cannot find any chiral primary operator with $R_0=3$ in the semi-classical analysis. Thus, it is natural to conjecture that there exist extra SUSY-current multiplets in the IR. 
Another supporting fact is that there is an exact match between the central charges of $U(1)_R$ and $U(1)_A$ which can be computed using localization \cite{Closset:2012vg,Closset:2012ru,Closset:2017zgf,Gang:2019jut}. This is expected if the symmetry enhancement occurs, as they would be related by an element of the Weyl group of $SO(4)_R$. Finally, the $\CT_r$ theory has a dual field theory description with manifest $\CN=3$ SUSY \cite{DG2023-2}. Combining the manifest $\CN=3$ symmetry with the superconformal index computation, one can argue that the symmetry is enhanced \cite{Evtikhiev:2017heo}. 

\section{Two topological twists}

Being 3d $\CN=4$ theories, each of the IR SCFTs admits two nilpotent topological supercharges $Q_A$ and $Q_B$, which we can use to perform two topological twists. They are defined by replacing the $SU(2)_E$ rotation group by the diagonally embedded $SU(2)$ subgroup of $SU(2)_E \times SU(2)_H$ or $SU(2)_E \times SU(2)_C$, which we call the topological A-twist or B-twist. We denote the resulting topological field theories by TFT$_A$ and TFT$_B$ respectively.

The local operators of the two topologically twisted theories are the Coulomb branch chiral rings and the Higgs branch chiral rings respectively. At the level of the superconformal index, the topological A-twist is realized by taking the limit $\nu \rightarrow -1$ with $\eta =1$, while the topological B-twist is realized by taking the limit $\nu\rightarrow 1$ with $\eta =1$ \cite{Razamat:2014pta,Bullimore:2020jdq}. For the class of theories discussed in the previous section, we find
 \begin{align}
 \CI_{\rm sci} (q, \eta=1, \nu = \pm 1) = 1\ ,
 \end{align}
which agrees with the expectation that the Higgs and Coulomb branches are trivial for this class of theories. 

In this paper, we focus on the properties of the A-twisted theories and leave a general analysis for the B-twisted theories in an upcoming paper by one of the authors \cite{GFK}.

\section{Fermionic sum representations of the minimal model characters}

The motivation for the UV description $\CT_r$ comes from the following expressions for the characters of non-unitary Virasoro minimal models $M(2,2r+3)$ \cite{Kedem:1993ze,Berkovich:1994es,Nahm:1994vas,Nahm:2004ch,welsh2005fermionic}:
\begin{align}
\begin{split}
	&\chi^{M(2,2r+3)}_{\alpha =0, \ldots, r} (q) 
	\\
	&=\sum_{m \in (\mathbb{Z}_{\geq 0})^r } \frac{q^{\frac{1}2 m^t K_r m +\sum_{a=1}^r a m_a - (Q_\alpha)^t m+h_\alpha- \frac{c}{24}}}{(q)_{m_1}\ldots (q)_{m_r}}\;, \label{characters of M(2,2r+3)}
\end{split}
\end{align}
where the $r\times r$ matrix $K_r$ coincides with the Chern-simons level matrix \eqref{level} and $Q_\alpha$ are rank-$r$ vectors whose components are
\begin{align}
\begin{split}
& (Q_\alpha)_{a} = \begin{cases}
		&0, \quad \alpha=0
\\
&\frac{1}2 (K_r)_{\alpha a}, \quad 1\leq \alpha \leq r
\end{cases}\;, \label{Qalpha}
\end{split}
\end{align}
We also define
\be
h_\alpha = \frac{\a (\a - 2 r-1)}{4r +6 } \textrm{ and } c = -\frac{2 r (6 r+5)}{2 r+3}\ ,
\ee
which are the conformal dimensions and the central charge.
Finally, the denominator is a product of $q$-Pochhammer symbols, defined by
\be
(q)_m = \prod_{i=1}^m (1-q^i)\ .
\ee
The simplest non-trivial example is the Virasoro minimal model $M(2,5)$, whose characters are
\be
\chi_0(q) = \sum_{m=0}^\infty \frac{q^{m^2+m+\frac{11}{60}}}{(q)_m}\ ,\quad \chi_1(q) = \sum_{m=0}^\infty \frac{q^{m^2 - \frac{1}{60}}}{(q)_m}\ .
\ee

These characters transform as a vector-valued modular function under the $SL(2,\mathbb{Z})$ transformation \eqref{modular transformation}.
\be
\chi_\alpha(-1/\tau) = \sum_{\beta} S_{\alpha\beta}\chi_\beta (\tau)\ ,\chi_\alpha(\tau+1) = \sum_{\beta} T_{\alpha\beta}\chi_\beta(\tau)\ ,
\ee
where $S$ and $T$ are the generators of $SL(2,\mathbb{Z})$ that satisfy the relation $S^2 = (ST)^3=I$ \footnote{In this letter, we only consider RCFTs with a trivial charge-conjugation matrix, i.e. $S^2=I$.}. They can be explicitly written as
\begin{align}
\begin{split}
&S_{\alpha \beta} = \frac{2(-1)^{r+\alpha +\beta}}{\sqrt{2r+3}}  \sin \left( \frac{2 \pi (\a+1)(\b+1)}{2r+3} \right)\;,
\\
&T_{\alpha \beta} = \delta_{\alpha, \beta} \exp \left(2\pi i (h_\alpha - \frac{c}{24})\right)\;.
\end{split} \label{modular}
\end{align}

We note that these are the simplest examples of characters that can be written in a so-called \emph{fermionic sum representation},
\be\label{fermionic sum}
\chi_{(A,B,C)}(q) = \sum_{m=(m_1,\cdots, m_r)\in (\mathbb{Z}_{\geq 0})^r} \frac{q^{\frac12 m^t A m + B^t m + C }}{(q)_{m_1}\cdots (q)_{m_r}}\ ,
\ee
where $A$ is a $r\times r$ positive definite symmetric matrix, $B$ is a $r$-dimensional vector and $C$ is a real number. 
There exists a large class of CFTs whose characters can be written in a fermionic sum representation or its generalizations. In particular, they have been extensively studied with regard to the characters of Virasoro minimal models \cite{Kedem:1993ze,Berkovich:1994es,Nahm:1994vas,Nahm:2004ch,welsh2005fermionic} and the characters of a certain class of logarithmic CFTs \cite{Feigin:2007sp}.

\section{Half-indices}

The half-index or the supersymmetric partition function on $D^2\times_q S^1$, introduced in \cite{Gadde:2013wq,Gadde:2013sca}, counts the boundary operators annihilated by the supercharges that are preserved by a chosen supersymmetric boundary condition on $\partial(D^2\times_q S^1) \simeq T^2$. More precisely, it computes
\be
I_{\text{half}} = \text{Tr}_{T^2} (-1)^{R_\nu} q^{\frac{R_\nu}{2} + j_3} \eta^A\ ,
\ee
where the trace counts the local operators on the boundary torus. If the boundary condition is compatible with the topological supercharge $Q_A$ (or $Q_B$) in the IR, the half-indices in the limit $\nu\rightarrow -1$ (or $\nu\rightarrow 1$) with $\eta =1$ calculate the characters of the boundary algebra for each topologically twisted theory.

The UV description of the $\CT_r$ theory \eqref{Tr UV} is designed in a way that its half-index reproduces the characters of the Virasoro minimal model $M(2,2r+3)$ in a specific limit. Indeed, if we impose the Dirichlet boundary conditions for all the $\CN=2$ $U(1)$ vector multiplets and the deformed Dirichlet boundary conditions for all the chiral multiplets in the $\CT_r$ theory \footnote{The deformed Dirichlet boundary condition is $\Phi|_{\partial}=c$ with non-zero $c$.}, the half-index reads \cite{Dimofte:2017tpi}\footnote{The peculiar shift of $\nu$ is required for $R_{\nu=0}$ to be the superconformal R-charge. We assign R-charges of zero to the chiral fields, which is compatible with the deformed Dirichlet boundary condition.}
\bea \label{half-index}
I_{\text{half}}(q,\eta,\nu) = &\sum_{m \in \mathbb{Z}^r} \frac{q^{\frac12 m^t K_rm}}{(q)_\infty^r} [(-q^{1/2})^{\nu-1}\eta]^{\sum_{a=1}^r a m_a} \\
&\times  \prod_{a=1}^r (q^{1-m_a};q)_\infty \;,
\eea
where we define $(x; q)_{\infty} := \prod_{n=0}^{\infty} (1-q^n x)$. 
We observe that this expression in the A-twist limit $ \eta=1, \nu=- 1$ coincides with the vacuum character 
of the Virasoro minimal model $M(2,2r+3)$ up to an overall factor \footnote{This observation was also made in recent work \cite{Jockers:2021omw}.},
\be
 \chi_{\alpha=0}^{M(2,2r+3)}(q) = q^{-\frac{c}{24}}I_\text{half}(q,1,-1) \ .
\ee
The characters of other modules $M_\alpha$ can be obtained by inserting loop operators. We consider the Wilson loops $L_{\alpha=1,\cdots, r}$, whose charge under the $a$-th $U(1)$ gauge group factor is given by the formula $(Q_\alpha)_a$ in \eqref{Qalpha}. The half-index $I_A[L_\alpha]$ in the presence of $L_\alpha$ reproduces the rest of the characters of $M(2,2r+3)$ \footnote{Wilson loop with gauge charge $\{Q_a\}_{a=1}^r$ contributes   a multiplicative factor $q^{\sum_a Q_a m_a}$  to  the summation in \eqref{half-index}.}
\be
\chi_{\alpha}^{M(2,2r+3)}(q) = q^{h_\alpha - \frac{c}{24}} I_A[L_\alpha](q) \ ,
\ee
for all $\alpha=1,\cdots, r$.
The choice of these particular sets of loop operators will be justified in the following section. 

In order to claim that these expressions are the characters of the boundary algebra of TFT$_A$ it is crucial to ensure that the boundary conditions are compatible with the topological supercharge $Q_A$ in the IR theory. In general this is a non-trivial task for theories which only have $\CN=2$ descriptions. See \cite{GFK} for the discussion of this issue in the context of deformable boundary conditions in the holomorphic-topologically twisted theory.

\section{Partition Functions on Seifert manifolds}

The supersymmetric partition functions of $\CN=2$  theories on a Seifert three-manifold $\CM_3$ are completely determined by the twisted effective superpotential $W$ and the dilaton potential $\Omega$ \cite{Nekrasov:2009uh,Nekrasov:2014xaa,Closset:2017zgf,Closset:2018ghr}. For the $\CT_r$ theory, we have 
\bea \label{W and Omega}
W_r(u) =& \sum_{a,b=1}^r\frac12 (K_r)_{ab} u_a u_b + \sum_{a=1}^r \zeta a u_a \\
&+\frac{1}{(2\pi i)^2} \sum_{a=1}^r\text{Li}_2 (e^{2\pi i u_a})\ ,\\
\Omega_r(u) =& \sum_{a=1}^r \frac{1}{2\pi i}\log(1-e^{2\pi i u_a}) + (\nu-1)a u_a  \ ,
\eea
where $\zeta$ is the real mass parameter for $U(1)_A$ symmetry. 
If $\CM_3$ is a degree-$p$ circle bundle over a genus $g$ Riemann surface, the partition functions can be written as
\be\label{partition function}
Z_{g,p}[\CT_r] = \sum_{\{P(u^*)=1\}} \CH^{g-1}(u^*) \CF^p(u^*)\ ,
\ee
where 
\bea\label{H and F}
\CH(u) = & \exp[2\pi i \Omega_r] \det_{ab} \partial_a\partial_b W_r\ ,\\
\CF(u) = & \exp[2\pi i (W_r - u_a \partial_a W  -\zeta \partial_\zeta W)]
\eea
with $x_a = e^{2\pi i u_a}$ \footnote{For F-maximization on a round sphere ($p=1,g=0$), we must perform a large gauge transformation to the background with zero R-symmetry flux on $S^2$, as discussed in \cite{Closset:2017zgf}. This can be done by shifting $\zeta \rightarrow \zeta + (\nu-1)$ and $\Omega \rightarrow \Omega -(\nu-1)\partial_\zeta W$ in \eqref{W and Omega} and \eqref{H and F}}.
These functions are then evaluated on the solutions to the so-called Bethe equations:
\be
P(u)=\exp\left[2\pi i \frac{\partial W_r(u) }{\partial u_a}\right] =1\ ,~~\text{for }a=1,\cdots,r\ ,
\ee
which reads, for $\CT_r$, 
\be
1-x_a = 
\eta^{a} \prod_{b=1} x_b^{(K_r)_{ab}} \ , ~~\eta = e^{2\pi i \zeta}\ .
\ee
This system of equations has exactly $r+1$ solutions, which we denote by $\{u^*_{\alpha=0,\cdots, r}\}$. 

In the twisting limit $(\nu, \eta)=(-1,1)$, the supersymmetric partition function \eqref{partition function} can  be written in terms of the modular data \footnote{The partition function \eqref{partition function} is computed in the background that is topologically twisted with the $U(1)_R$ symmetry of $\CN=2$ algebra. While this computation correctly reproduces the $S_{0\alpha}$, one has to argue that the same is true for the identification of $\CF$ and $T^{-1}_{\alpha\alpha}$. We simply observe here that $\CF(u_\alpha^*)$ computed in this background coincides with $T_{\alpha\alpha}$ for the half-indices of $\CT_r$'s.}:
\be \label{TQFT ptn}
Z_{g,p}[\CT_r]\big{|}_{(\nu, \eta)=(-1,1)} = \sum_{\alpha} S_{0\alpha}^{2-2g} T_{\alpha\alpha}^{-p}\ , 
\ee
where $\alpha$ labels  modules of boundary  algebra and  $(S,T)$ are the modular matrices that transform the characters as in \eqref{modular}. By comparing \eqref{partition function} and \eqref{TQFT ptn}, we can extract the modular data, more precisely the set $\{S_{0\alpha}^2,T^{-1}_{\alpha\alpha}\}$, by identifying it with $\{\CH(u_\alpha^*)^{-1}, \CF(u_\alpha^*)\}$ \footnote{The SUSY partition functions in \eqref{partition function} depend on various local counter-terms such as background CS level of R-symmetry as well as a choice of  3-manifold framing. They affect the overall phase factor of the partition function in the twisting limits. We do not keep track of all these subtle choices and the $T$-matrix is determined only up to an overall phase factor.}. 

The full modular data and the precise map between the Bethe vacua $u_\alpha^*$ and the modules $M_\alpha$ (or equivalently, the loop operators $L_\alpha$) can be constructed by requiring the relation \cite{Cho:2020ljj}\footnote{ The Bethe-vacuum $u_{\alpha=0}^*$  corresponding to the vacuum module is chosen to  satisfy $|Z_{g=0, p=1} \textrm{ in  \eqref{partition function}}|=\CH(u_{\alpha=0}^*)^{-1/2}$.}
\begin{align}
&L_\alpha (u_0^*) = S_{\alpha 0}/S_{0 0} = \pm \sqrt{\CH(u^*_0)/\CH(u^*_\alpha) }\ ,
\end{align}
where $L_\alpha (u^*_\beta)$ is the loop operator $L_\alpha$ evaluated on the Bethe-vacuum $u^*_\beta$. For this class of theories, we consider Wilson loops $L$ with gauge charges $(Q_1, \ldots, Q_r)$ which contributes
\begin{align}
L (u_\beta^*) = \prod (x_a^{-Q_a})|_{u \rightarrow u^*_\beta}\;.
\end{align}
Then the following identity 
\begin{align}
 S_{\alpha\beta}= L_\alpha (u_\beta^*) S_{0\beta} \textrm{ with } S_{0\beta}=  \pm 1/ \sqrt{\CH(u_{\beta}^*)}\;,
\end{align}
together with the $SL(2,\mathbb{Z})$ relations and the non-negativity of the fusion rule coefficients, determines the $S$ matrix up to an overall sign and the $T$ matrix up to an overall phase factor of the form $\exp(\frac{2\pi i\mathbb{Z}}3)$. This procedure determines the precise set of lines $L_\alpha$ as stated in the previous section. The $S$- and $T$-matrices computed in this way agree with the modular matrices of the Virasoro minimal model $M(2,2r+3)$, as given in \eqref{modular}.

\section{Discussion}

In this letter, we constructed a class of rank-0 SCFTs inspired by the characters of non-unitary minimal models. By applying the correspondence in the reverse direction, a novel class of non-unitary RCFTs is introduced \cite{DG2023}, which correspond to well-studied examples of rank-0 SCFTs. In this way, we expect the correspondence to help explore uncharted landscapes of 2d RCFTs as well as 3d SCFTs.

The fermionic sum formulas for characters are known for a much larger class of RCFTs. In particular, Nahm \cite{Nahm:2004ch} and Zagier \cite{Zagier:2007knq} classified modular functions of the form of \eqref{fermionic sum} for small values of $r$, which can therefore be candidates for characters of an RCFT. In an upcoming work \cite{GKS}, we will extensively study the classification of the rank-0 theories that give rise to these characters and discuss the relation to the work of Zagier \cite{Zagier:2007knq}.

One of the important questions is whether one can explicitly construct the boundary rational vertex algebras for the TFTs discussed in this paper. As a first step toward this goal, in an upcoming paper \cite{GFK} by one of the authors, the boundary algebras for the B-twisted theories will be studied via the holomorphic-topological twist of the UV gauge theories.

\begin{acknowledgments}
We wish to acknowledge Andrea Ferrari, Niklas Garner, Dongyeob Kim, Sungjay Lee, Jaewon Song for discussions and collaborations on related works. The work of DG is supported in part by the National Research Foundation of Korea grant NRF-2022R1C1C1011979. The work of HK is supported by the Ministry of Education of the Republic of Korea and the National Research Foundation of Korea grant NRF-2023R1A2C1004965.  The work of SS is supported by the US Department of Energy under grant DE-SC0010008.
\end{acknowledgments}

\appendix

\bibliography{minimal_rank0}% Produces the bibliography via BibTeX.

\end{document}